\documentclass[fleqn,12pt,twoside]{article}
\usepackage{espcrc1}

\usepackage{graphicx}

\newcommand{\ak}{a_\kvec^{\phantom{\dagger}}}
\newcommand{\akdag}{a_\kvec^{\dagger}}
\newcommand{\beq}{\begin{equation}}
\newcommand{\eeq}{\end{equation}}
\newcommand{\beqa}{\begin{eqnarray}}
\newcommand{\eeqa}{\end{eqnarray}}

\newcommand{\kf}{k_{\scriptscriptstyle\rm F}}
\newcommand{\kvec}{{\bf k}}

\title{Three-Body Interactions in Many-Body Effective Field Theory}

\author{R. J. Furnstahl\address{Dept.\ of Physics, Ohio State University, 
         Columbus, OH 43210, 
         USA\\E-mail: furnstahl.1@osu.edu}%
         \thanks{This work was supported in part by the U.S.\ NSF
           under Grant No.\ PHY-0098645.}}

\begin{document}

\maketitle

\maketitle

\begin{abstract}
This contribution is 
an advertisement for applying effective field
theory (EFT) to many-body problems, including nuclei and cold atomic
gases.
Examples involving three-body interactions
are used to illustrate how EFT's
quantify and systematically eliminate
model dependence, and how they make many-body 
calculations simpler and more powerful. 
\end{abstract}

\section{Introduction}

A general principle of \emph{any}
effective low-energy theory is that if a system
is probed or interacts at low energies, resolution is also low, and fine
details of what happens at short distances or in high-energy
intermediate states are not resolved \cite{CROSSING}.
In this case, it is easier and more efficient to use low-energy degrees
of freedom for low-energy processes.
The short-distance structure can be replaced by something simpler
(and wrong at short distances!) without distorting low-energy
observables.
There are many ways to replace the structure; an illuminating way is
to lower a cutoff $\Lambda$ on intermediate states.

\begin{figure}[t] 
 \begin{minipage}{2.9in}
   \includegraphics[width=2.85in,angle=0]{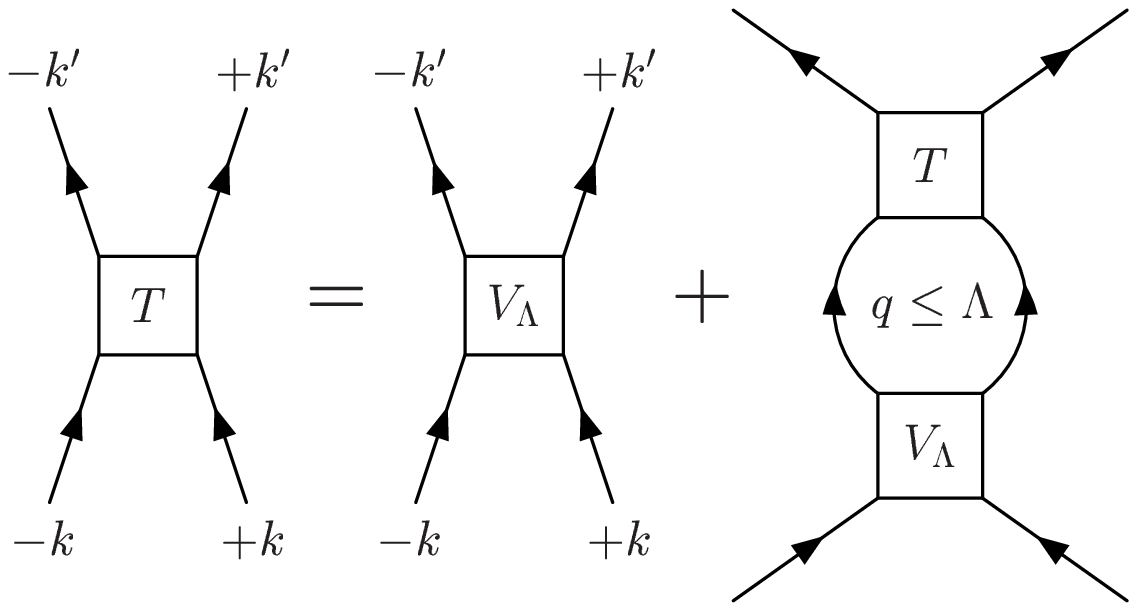}
   \vspace*{.3in}
   \centerline{%
     \includegraphics[width=2.0in,angle=0]{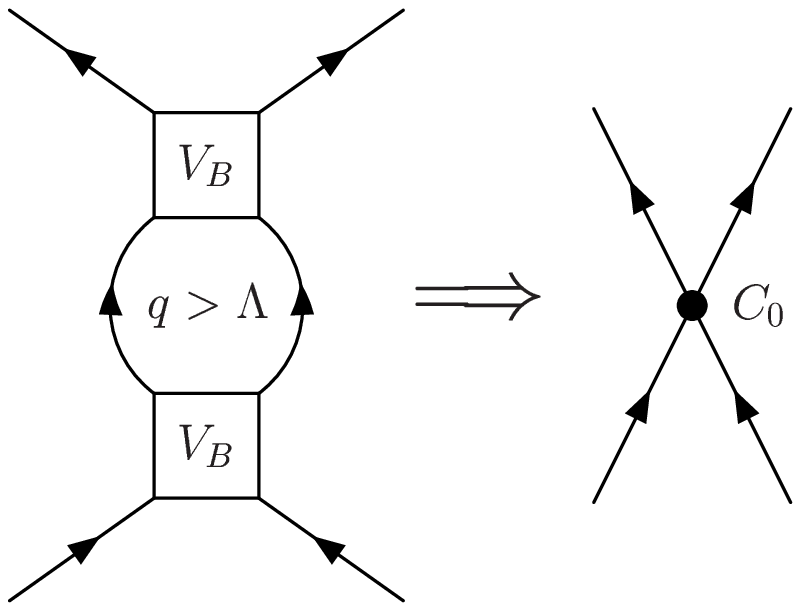}%
   }
 \end{minipage}
 \hfill
 \begin{minipage}{3.1in}
   \vspace*{-.2in}
   \includegraphics*[width=3.0in]{all7bw}
 \end{minipage}
 \vspace*{-.3in}
 \caption{Left: Equation for the $T$-matrix
   with cut-off potential $V_\Lambda$ and
   replacement of the effects of high $q$ states with a contact interaction. 
   Right: $\chi^2/\mbox{dof}\approx 1$  potentials in $^1S_0$
   channel and corresponding 
   $V_\Lambda(k,k)$'s at $\Lambda=2.1\,\mbox{fm}^{-1}$ (black dots)
    \cite{SCHWENK03}.}
  \label{fig:one}
\end{figure}

Consider nucleon-nucleon scattering
in the center-of-mass frame (see Fig.~\ref{fig:one}).  
The Lippmann-Schwinger equation iterates
a potential that we take originally as one of the $\chi^2/\mbox{dof}\approx
1$ potentials.  Intermediate states with relative momenta
as high as $q \approx 20\,\mbox{fm}^{-1}$
may be needed for convergence.  Yet the data and the reliable long-distance
physics (pion exchange)
only constrain the potential for $q \leq 3\,\mbox{fm}^{-1}$.
We can cut off the intermediate states at successively
lower $\Lambda$; with each step we
have to change the potential $V_\Lambda$ to maintain the same
phase shifts.
This determines a renormalization group (RG) equation
for $V_\Lambda$  \cite{SCHWENK03}.
We see in Fig.~\ref{fig:one} that at $\Lambda=2.1\,\mbox{fm}^{-1}$, 
corresponding to 350\,MeV lab energy,
the
potentials have all collapsed to the same low-momentum potential
(``$V_{{\rm low\,}k}$'').
The net shifts are largely constant in momentum
space, which means they are well represented by
contact terms and a derivative expansion.  This observation suggests a local 
Lagrangian approach.
We also note that the high-momentum
$\Lambda$ dependence for two-nucleon
scattering appears as powers of $\Lambda$ only (no
logarithms).

The low-energy data is insensitive to \emph{details} of short-distance
physics, so we can replace the latter with something simpler without
distorting the low-energy physics.  Effective field theory (EFT) is a
local Lagrangian, model-independent approach to this program.
Complete sets of operators at each order in an expansion permit
systematic calculations with well-defined power counting.
The program is realized as described in Ref.~\cite{CROSSING}, which we
apply to a basic many-body system, the dilute Fermi gas:
\begin{enumerate}
 \item \emph{Use the most general Lagrangian with low-energy dof's consistent
 with global and local symmetries of the underlying theory.}  For a
 dilute Fermi system, this takes the form (with omitted derivative
 and higher many-body terms):
 \beq
  {\cal L}_{\rm eft} = 
               \psi^\dagger \bigl[i\frac{\partial}{\partial t} 
               + \frac{\nabla^{\,2}}{2M}\bigr]
                 \psi - \frac{{C_0}}{2}(\psi^\dagger \psi)^2
                    - \frac{{D_0}}{6}(\psi^\dagger \psi)^3 +  \ldots
 \eeq
 \item \emph{Declare a regularization and renormalization scheme.} For a
 natural scattering length $a_s$ (e.g., hard spheres where
 $a_s \propto R$, the sphere radius), 
 dimensional regularization and minimal
 subtraction (DR/MS) are particularly advantageous \cite{HAMMER00}.
 A simple matching to the effective range expansion for
 two-body scattering determines the two-body coefficients ($C_i$)
 to any desired order.  For example, $C_0 = 4\pi a_s/M$.
 \item \emph{Establish a well-defined power counting;} e.g., the energy
 density ${\cal E}$ in 
 powers of $\kf a_s$:
 
  \includegraphics*[width=3.8in]{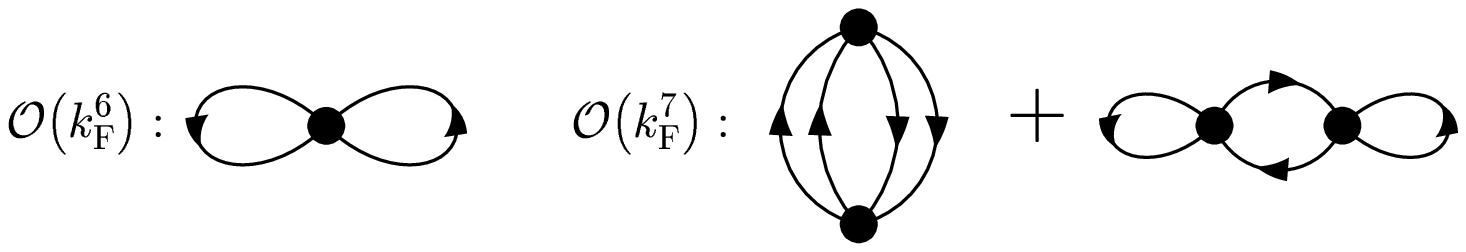}
  \raisebox{.3in}{\mbox{\quad and so on.  The rules yield}}
  \vspace*{-.3in}
  \beq
   {\cal E} =
      {\rho} \frac{{\kf^2}}{2M}
       \biggl[\frac{3}{5} + {\frac{2}{3\pi}}{(\kf a_s)}
       +
         {\frac{4}{35\pi^2}(11-2\ln 2)}{(\kf a_s)^2}
         + \cdots
       \biggr] \ .  
       \label{eq:edensity}
  \eeq      
\end{enumerate}
The calculation of the energy density is far easier in the EFT approach
than in conventional treatments \cite{HAMMER00}.
For example, each additional $C_0$ vertex 
simply brings a single power of $\kf a_s$. 
The contribution for each diagram is a coefficient with all of the
dependence on the short-range scale (e.g., $\Lambda \sim 1/R$) times a
multi-dimensional integral that is simply a geometric factor (and which
is conveniently evaluated even at high order using Monte Carlo
integration).

\section{Inevitability of Three-Body Interactions}

Naively, it would appear from (\ref{eq:edensity}) that the energy density
is a power series in $\kf a_s$.
In fact, the polynomial in $\kf a_s$ is disrupted by three-body
contributions. 
(The following contributions assume 
the spin/isospin degeneracy is greater than two.)
These emerge \emph{inevitably} in the form of logarithmic divergences in 3--to--3  
scattering (left two diagrams):

 \includegraphics*[width=3.5in,angle=0]{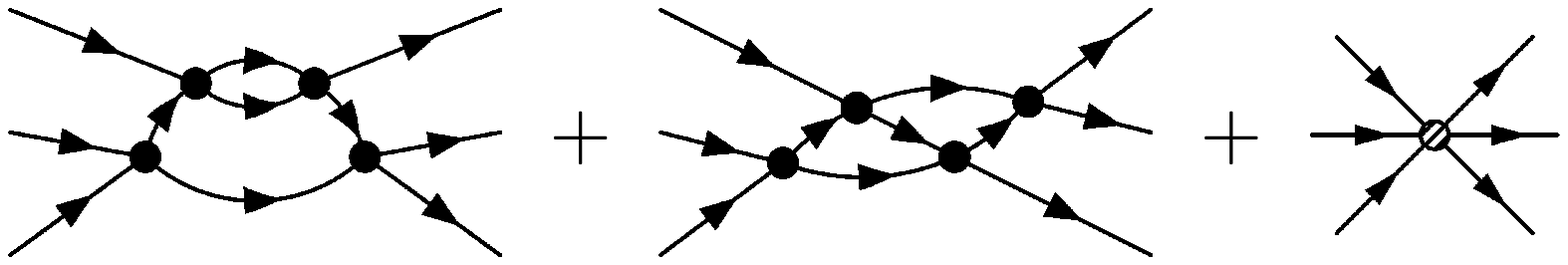}
  \quad\raisebox{.27in}{$\stackrel{{\displaystyle\cal E}}{\Longrightarrow}$}\quad
 \includegraphics*[width=2.0in,angle=0]{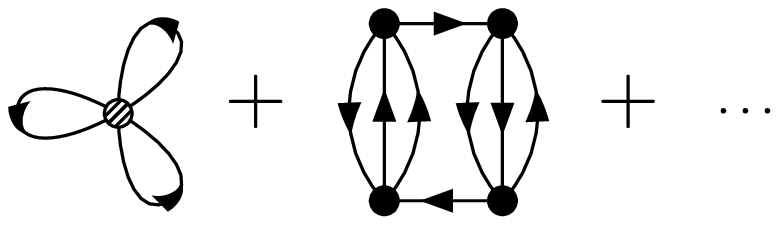}

\noindent
The divergence is easily isolated and in dimensional regularization 
the amplitude is
\beq
  {\cal T}_{3\to 3}^{\rm ln}
     = -iM^3 (C_0)^4 \,\frac{4\pi-3\sqrt{3}}{8\pi^3}
     \Bigl[ \frac{1}{D-3} - 2\,{\ln\mu} + \cdots \Bigr] \ .
\eeq
Changes in the parameter $\mu$ are absorbed by the three-body
coupling $D_0(\mu)$, 
yielding an RG equation that is easily solved
for the $\mu$ dependence of $D_0$ since $C_0$ is constant:
\beq 
  \mu\frac{d}{d\mu}D_0=M^3 (C_0)^4\, \frac{4\pi-3\sqrt{3}}{4\pi^3}
   \ \Longrightarrow\
   D_0(\mu)=D_0(1/a_s)+M^3 (C_0)^4 \,\frac{4\pi-3\sqrt{3}}{4\pi^3}\,
      {\ln(a_s \mu)}\,.
      \label{eq:four}
\eeq
The $\ln(\mu)$ dependence from $D_0(\mu)$ in the energy (fourth diagram)
must be
canceled, which tells us for free there is a term in the energy 
density proportional to $\ln(\kf a_s)$  with the \emph{same}
coefficient as the $\ln(a_s\mu)$ term
in Eq.~(\ref{eq:four}) [see Ref.~\cite{HAMMER00} for
the complete details].
While the logarithm is determined, $D_0(1/a_s)$ is not:  two-body data
alone is insufficient!

We can further exploit the general structure of the renormalization
group equations to identify additional logarithms (and powers of
logarithms)
\cite{BRAATEN,HAMMER00}.
The scale $\mu$ only appears in logarithms, which means that matching
$\Lambda$ dimensions in the RG equations is very restrictive.
The couplings have dimension $C_{2i} \sim 4\pi/M\Lambda^{2i+1}$ and
$D_{2i} \sim 4\pi/M\Lambda^{2i+4}$, so the RG equation for the 
coefficient $C_0(\mu)$ can only have one $C_0$ on the
right side, which in turns tells us to look for log divergences
in 2--to--2 diagrams with a single
$C_0$:
\beq
  \mu \frac{d}{d\mu}C_0 = a\, C_0 \ \Longrightarrow\
    \mbox{tree-level only; no log divergence} \ \Longrightarrow\ a = 0
    \ \Longrightarrow\ C_0 = \mbox{const.}
\eeq
For the three-body, no-derivative coefficient $D_0(\mu)$,
we reproduce the form found above:
\beqa 
    \mu \frac{d}{d\mu}D_0 &\!\!=\!\!& a\, (C_0)^4 
      + b\, C_0 C_2 + c\, C_0 C_2' + d D_0
     \ \Longrightarrow\  b=0,\ c=0,\ d=0\ \mbox{(3--to--3 tree level)} 
     \nonumber \\
  & & \Longrightarrow\  
  \frac{d D_0}{d\ln\mu} = a (C_0)^4
   \
     \Longrightarrow\ 
     D_0(\mu) = a (C_0)^4\ln\mu  + {const.}
\eeqa
If the right side has $D_0 \propto \ln\mu$, then the
coefficient goes like $(\ln\mu)^2$, and so on \cite{BRAATEN}.

\section{Observables}

\begin{figure}[ht]
  \centerline{%
    \includegraphics*[width=3.in,angle=0]{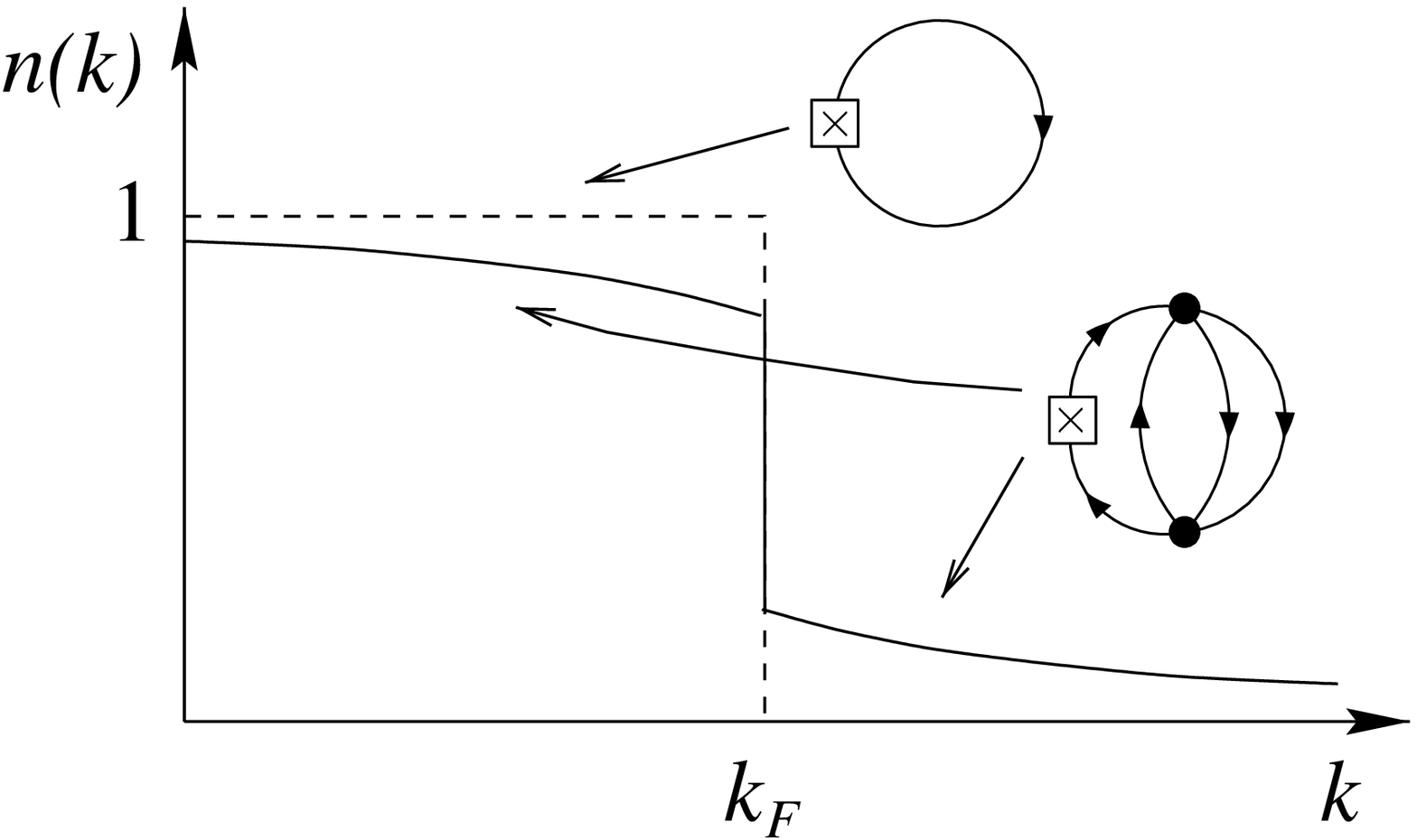}
    \hfill
    \raisebox{.2in}{%
    \includegraphics*[width=3.in,angle=0]{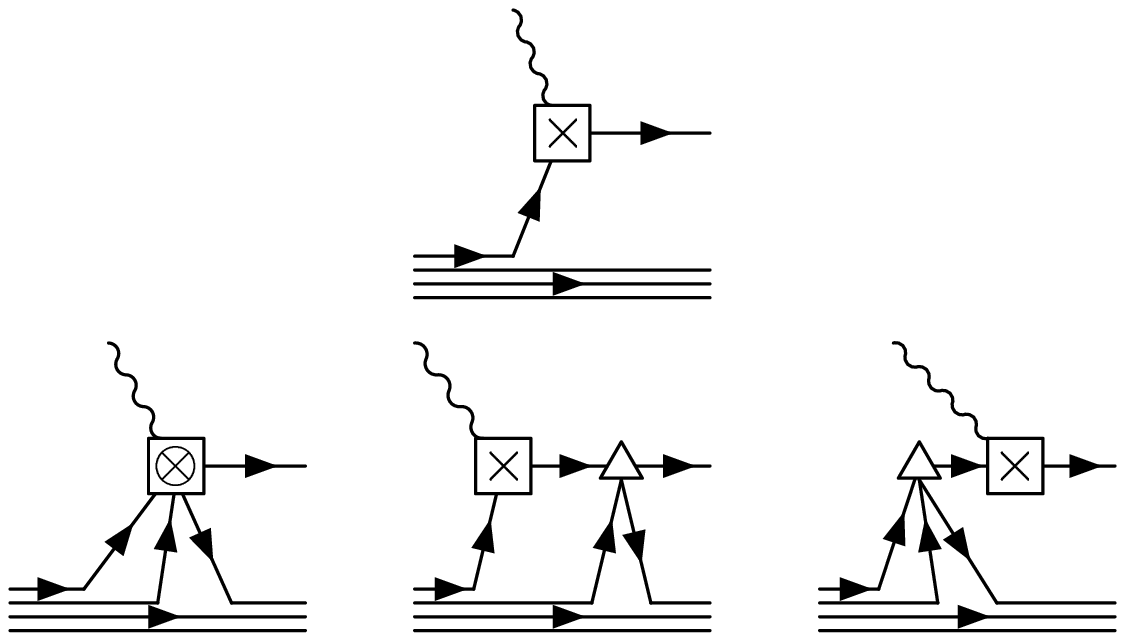}}  }
  \vspace*{-.3in}
  \caption{Left: Diagrammatic momentum distribution.  Right: Schematic
    $(e,e'p)$ process.}
  \label{fig:two}
\end{figure}

An example of how subtle model dependence is 
clearly identified by the EFT arises
when considering
occupation numbers, which are typically treated as many-body observables.
In a uniform system with second-quantized creation and destruction operators
$\akdag$ and $\ak$, the momentum (occupation) distribution is $n(k) =
\langle \akdag\ak\rangle$, which measures the strength of 
correlations (Fig.~\ref{fig:two}, left).
It is said to be measurable in $(e,e'p)$ on a nucleus.
But is $n(k)$ an observable?

\vspace*{-.3in}

\vspace*{-.1in}

The status of potential observables can be tested using local field
redefinitions, such as
$\psi \longrightarrow \psi +
       \alpha (4\pi/{\Lambda^3}) (\psi^{\dagger}\psi) \psi
$
with arbitrary $\alpha$; if
$\alpha \sim {\cal O}(1)$ this is ``natural''
 \cite{FURNSTAHL01}.
(These field redefinitions are analogous to, but not the same
as, unitary transformations.)
Such a redefinition induces both two-body off-shell vertices (triangles) 
\emph{and}
three-body vertices.
It can be shown that
the energy density is model  independent (i.e., independent of $\alpha$) 
\emph{if} all
terms are kept \cite{FURNSTAHL01}: \\  
\hspace*{1in}\includegraphics*[width=3.5in,angle=0]%
  {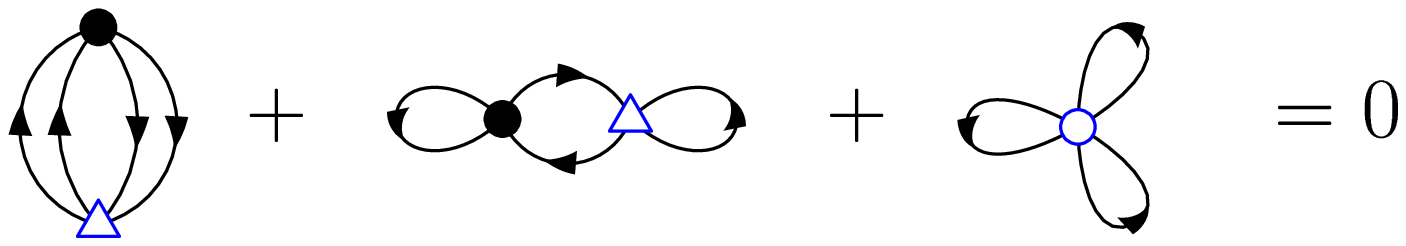}

\noindent
In this example, the one-body kinetic term generates the triangle vertex
under the redefinition while the two-body no-derivative term generates
the three-body vertex (open circle).
If the three-body terms are omitted, then the energy would depend on
$\alpha$ (even though the different forces reproduce the
same two-body phase shifts) and then one
might be fooled into thinking that $\alpha$ can be determined by
comparison to experiment.
The
energies for different $\alpha$'s would lie along a ``Coester line,''
which is just a form of model dependence 
(``off-shell ambiguities'') made manifest by the EFT
\cite{FURNSTAHL01}.

There are similar induced contributions 
to the momentum distribution, with the
additional issue that the corresponding operator is changed 
by redefinitions
and there is no preferred
definition (there is no Noether current, as for the
fermion number) \cite{FURNSTAHL02a}.
These induced contributions
correct the impulse approximation when analyzing $(e,e'p)$ experiments,
mixing vertex corrections (exchange currents) and initial and final
state interactions in an $\alpha$-dependent way
(see Fig.~\ref{fig:two} right).
This means that the distribution $n(k)$ is not directly accessible;
more generally, 
experiment cannot resolve ambiguities in momentum distributions within a
calculational framework based on low-energy degrees of freedom.
Instead the distributions
 are auxiliary quantities defined only in a specific
convention; they are useful within this convention but are not observables
(this is analogous to quark distributions in deep inelastic scattering).
However, the ambiguities have a natural size \cite{FURNSTAHL02a}; 
if they are negligible
then the momentum distributions are effectively observables.

\section{Current Trends in Many-Body EFT}

The EFT tools and techniques offer many new possibilities for the
systematic and model-independent calculation of many-body systems;
the examples here involving three-body interactions
are just a sampler.
When contributions to three- and higher-body scattering from multiple
short-distance two-body scatterings have
logarithmic divergences at large intermediate-state momentum,
they are not resolved and
three-body interactions \emph{must} be included to avoid model
dependence.
Careful consideration of the regulator dependence turns a necessity into a
virtue, providing valuable information about the analytic structure
of observables.
The second example illustrated how local field redefinitions
are a clean tool for assessing potential observables.
Simple rule:  if a calculated quantity depends on a transformation
parameter $\alpha$, it is either not an observable or you've forgotten
some contribution.  
These transformations also demonstrate explicitly how different two-body
forces are associated with different three-body forces.

Other topics under current investigation include nonperturbative EFT and
applications to finite systems.
The effective action formalism has been used for a nonperturbative
large N expansion in Ref.~\cite{FURNSTAHL02} and work is in progress
to extend the EFT approach to large $a_s$ that was initiated by Steele
\cite{STEELE00}.
A merger of density functional theory (DFT) and EFT is presented in
Ref.~\cite{PUGLIA03}, with on-going work on long-range forces,
pairing, and a systematic gradient expansion.
Some planned applications are energy functionals for nuclei far from
stability and superfluidity in trapped fermionic atoms.
Other groups are adapting chiral perturbation theory to many-body
systems \cite{CROSSING} 
and there is an EFT program for bosonic systems by Braaten, Hammer, 
and collaborators \cite{BRAATENB}.

\end{document}